# Distance Teaching Experience of Campus-based Teachers at Times of Pandemic Confinement

Abbas Cheddad and Christian Nordahl

Department of Computer Science, Blekinge Institute of Technology, 371 79, Karlskrona, Sweden.

## Abstract

Amidst the outbreak of the coronavirus (COVID-19) pandemic, distance education, where the learning process is conducted online, has become the norm. Campus-based programs and courses have been redesigned in a timely manner which was a challenge for teachers not used to distance teaching. Students' engagement and active participation become an issue; add to that new emerging effects associating with this set-up, such as the so called "Zoom fatigue", which was coined recently by some authors. In realising this problem, solutions were suggested in the literature to help trigger students' engagement and enhance teachers' experience in online teaching. This study analyses these effects along with our teachers' experience in the new learning environment and concludes by devising some recommendations. To attain the above objectives, we conducted online interviews with six of our teachers, transcribed the content of the videos and then applied the inductive research approach to assess the results.

## 1 Background

As we write this paper, hopefully, we are amidst the final stages of a global pandemic, during which the world has experienced unprecedented turbulence at all levels, including uncertainty in the academic upbringing [1] [2]. The student experience is an important concept, and looking after the student's academic well-being is only one part of the picture [3], especially during this time of pandemic confinement. Previous literature confirmed that working from home increases the flexibility of attaining academic tasks and correlates positively with overall task satisfaction. However, it can also lead to more stress and negative personal well-being, according to Anderson et al. [4]. In academia, Swedish universities have been more fortunate than institutions in other countries to transition our teaching and research swiftly and smoothly to be performed remotely to comply with the lockdown protocols. Based on the experiences of both teachers and students over the last year, we believe that things will not necessarily go back to exactly how they once were. For instance, course structures were altered, lectures reimagined, examination content and set-up redesigned, and so on. Even though they were performed due to necessity, some of these adaptations may have established interest in change in both teachers and students.

There is a plethora of literature discussing student engagement mainly in the traditional distance teaching settings [5, 6, 7], campus teaching [8, 9], or as a combination [10]. The term

"*Engagement*" refers to the level of attention, interest, optimism, and passion students show when taught a given subject. This can be reflected through notably asking/answering questions and interacting with the teacher, for example. With the sudden change to online teaching, it is reasonable to believe that many elements were ported into the online setting. The courses might not have been altered to fully facilitate the distance teaching setting, as there was not enough time. Additionally, the courses taught were not intended to stay online forever either. The goal of the universities have remained to continue campus teaching as usual when the pandemic has passed.

Some recent publications deal with the COVID-19 situation but mainly in the challenges that it imposes [11], in the ways to increase productivity for businesses [2], in scrutinizing work-life conflict during pandemics [12], or in providing a general diagnosis [13]. One of the major discussion points from the literature has brought up the issue of interaction with students during this pandemic. With most students having cameras turned off, giving lectures online does not allow interaction with the students. It is harder to judge their engagement as compared to seeing individual reaction and interaction of students during traditional lectures. It is a challenge to gauge the level of understanding of lectured topics if we lack such non-verbal interactions.

# 2 Research Problem

In the realm of the COVID-19 pandemic and its consequences on academia, it would be interesting to investigate what our teachers have identified as issues during remote teaching and what they have learned from such experience. There are a few of papers that try to capture teachers' emerging experience out of this sudden shift to remote teaching and their adaptability [14, 15]. In this study, we want to contemplate the same matter among our teachers at the Department of Computer Science (DIDA) at Blekinge Institute of Technology (BTH) and we enrich that with self-reflection (which is an integral part of the process by which individuals in higher education become more critical beings [16]). Hence, this study sheds lights into our teachers' experience and their recommendations when it comes to online teaching tailored to a pandemic and lock-down scenario. We aim to identify what common issues have been seen by the teachers, what they have adopted and, foremost, what they have learned.

## 2.1 Research Questions

**RQ1:** *How did our teachers adapt to distance teaching within Computer Science at this time of pandemic and lock-down?*

*Motivation:* Given the rapid paradigm shift in education due to the COVID- 19 pandemic, most, if not all, courses which are given by DIDA (BTH) had to be adopted to a distance-based format in a very short time frame. During this time, aids were provided on how to convert and structure our courses successfully—for example, the work of our colleagues Eriksson et al. [17].

**RQ2:** *What are the possible ways that our teachers perceive to get the learning platform engaging?*

*Motivation:* The courses that are being taught currently amid this pandemic are in their core campus-based courses. Teaching them online is not analogous to teaching courses that are originally designed as distance courses. This pandemic situation mandates distance teaching, which brought a unique experience to our teachers different from a campus-based teaching set-up. In this sense, there might be some positive elements in this experience that could benefit the campus-based teaching when our university re-opens for that.

**RQ3:** *Do our teachers believe that their new teaching experience (distance teaching) could beneflt their campus-based teaching in the future?*

*Motivation:* Less focus is put on explicitly keeping both students and teachers *engaged* during the pandemic and its mandatory physical distancing. In the current situation, all courses are given by distance-teaching (e.g., home-learning and home-examination). Therefore, this study is highly relevant and correlates strongly with our collective aim to ameliorate the quality of distance learning, which our institute is relatively new to.

## 2.2 Objectives

To answer the questions stated in the previous section, we divide them into the following objectives:

- To identify the current landscape and terminology from the literature regarding distance-based learning during times of physical distancing.

  *Motivation:* Many of the prior work relate to challenges and issues facing distance-learning during pandemics including a few studies that deal with boosting productivity in traditional learning scenarios (i.e., campus-based). Studying such literature may give us hints and/or help us frame a well thought set of contemplations to discuss with our teachers on how they experience distance learning during lockdown times.

- To determine teachers' point of view on the adaptations made during the COVID-19 pandemic.

  *Motivation:* Without listening to teachers and getting their point of view and valuable inputs, one may produce incomplete and/or biased diagnosis and thereafter unfruitful recommendations.

- To draw some recommendations built upon scrutinising the interviewees' personal experiences to help increase students' engagement.

  *Motivation:* The final recommendations that this study aims for form the core outcome. There is a lack of such guidelines and recommendations to increase students' engagement during this challenging time we are experiencing and which we do not know for sure when it will end.

# 3 Related Work / State of Knowledge

Amid the COVID-19 pandemic, the online teaching scenario becomes no longer an alternative, teachers within the traditional learning (campus based) platform found themselves suddenly required to adopt distance teaching. A modest number of literature papers discuss the current lockdown situation and its impact on the learning atmosphere including ways to get the best out of the distance learning technology. The effect of the COVID-19 pandemic has been investigated in a few different ways. A study shedding light upon the psychological well-being of students in China [18], showed that a great deal of anxiety was felt during the initial months of the pandemic, especially for students with relatives that had gotten ill.

Another study was performed in Pakistan, where they investigated the student perspective of the e-learning adaptation of COVID-19 [19]. Due to a lack of well-functioning Internet infrastructure, many of the students had gone through struggles in performing their academic activities. Furthermore, projects, seminars, and other group activities were perceived as harder to conduct than before. Similarly, a study about the effect of a poor Internet infrastructure was performed in Indonesia [20].

It has been proven harder to both transition from a traditional set of classroom lectures into an online format as well as monitoring the progress of students [21]. Students perceived positively the new nature of e-learning partly due to it being in the asynchronous mode, according to literature [22]. Now they are able to access the information in their own time, instead of being forced to attend lectures.

On the other side of the spectrum, the sudden shift to online learning has shown some difficulties for the students in the adaptation. Specifically, the importance of time management and ability to focus for a long period of time behind the computer screen [23]. Along the same line, Wiederhold [24] published a short editorial communication letter in which she discusses what it becomes known as the "Zoom fatigue". She brings about the mental exertion, lack of nonverbal cues, whether others on the call are with the teacher and following, whether the message went through, etc. Video calls leave students and teachers alike feeling unsettled and exhausted. To increase students engagement and enhancing non-verbal communication, a suggestion is brought forward to shift from Zoom to other platforms that allow learning through virtual reality (VR) with avatars, needless to say, such suggestion requires special equipment (i.e., VR headset and associated technology) which are not attainable by the majority of the students (e.g., cost restrictions). Realising this constraint, Wiederhold [24] suggested an alternative to break down the "Zoom fatigue" by giving a set of best practices, such as to look at the camera, to limit use of videoconferencing technology by trying to staggering meetings with non-screen breaks in between [25, 24]. AbuJarour et al. [26] investigated factors that could impact academics' productivity while working from home during the pandemic lockdown. The study uses a survey conducted at different parts of the world. Their results show that both personal and technology-related factors impact productivity.

# 4 Methodology

Notably, there are many important advantages of e-learning and its effective role in disseminating knowledge, however, overcoming space and time restrictions in the educational process to help increase students engagement is a challenge especially when e-learning is no longer merely an option but the only option. To answer the research questions, this study will be based on open-ended interviews via Zoom with predefined focused group (i.e., teachers) from our Computer Science department at BTH. Whether our learning environment is set-up in a synchronized or asynchronous manner, it must embody the spirit of flexible academic orientation well, especially when wedded to an unenviable pandemic reality. There is no clear strategy in place and the academic atmosphere in Sweden is lurching from one crisis to another in dealing optimally with students' engagement during this pandemic time.

Our interviewees were made aware of the purpose of their participation and the way we intend to collect and analyse the data (i.e., recorded meeting to be transcribed), therefore, their initial consent was needed. The issues to be discussed were to be sent to them in a good time prior to having the actual interview, giving them ample time to reflect upon the raised matters.

To investigate the uncharted side of the collected transcribed data from the interviews, we adopted the qualitative analysis approach.

## 4.1 Participating Teachers

Teachers at the department of Computer Science at BTH were contacted by email to explain the motive behind our intended interview and the protocol and to seek their consent to participate. Issues/questions to be discussed (the contemplation points, see Section 4.4) were sent to the interviewees some days prior to the actual interview. The interview was conducted online via Zoom and each session lasted for approximately one hour for each teacher. The interview took place end of March/beginning of April 2021. Participating teachers were informed that the interview will be recorded for the sake of post-transcription and for comparison between participants' inputs. The participants had the option to revoke their permission to use their interview if we were notified 24h post-interview. As for the confidentiality of records, the data was stripped out from any confidential information (i.e., the transcribed text is anonymized). The interview ended up with 6 participants, 3 male and 3 female teachers from different disciplines, see Table. 1.

Table 1: Interview participants from the Computer Science department.

| Participant | Gender | Title | Teaching Area |
|---|---|---|---|
| 1 | M | Dr. | Computer Security |
| 2 | M | Dr. | Data Science |
| 3 | F | Dr. Docent | Digital Scheduling |
| 4 | F | Prof. | Software Engineering |
| 5 | F | Dr. Docent | Game Technology |
| 6 | M | Prof. | Telecommunication |

## 4.2 In-Depth Qualitative Analysis

Perhaps, the most intuitive analysis for interviews is the qualitative approach since we are after perception, opinion, reflections, etc. which cannot be quantified. We intentionally avoided questionnaires since they limit the outcome to our own perception and because "... responsive interviewing is generally gentle and cooperative, feels respectful and is ethical" [27]. Thus, we value the fact that "the interviewer must nevertheless recognize that the meaning is, to some degree, a function of the participant's interaction with the interviewer", according to Seidman [28].

**Coding and thematic analysis:** Interviews were transcribed from their initial state (i.e., video recording of an oral interview). Coding (content analysis), labeling the categories (organizing the code into themes) and subsequently describing the connections between them, will eventually form the main addendum which connects the dots and unravels any findings.

The naturalist and interpretive constructionist perspectives model was followed since we were concerned with the goggles through which our interviewees viewed and assigned meanings to events around them and how they interpreted them. In this model, we, as researchers supervising this study need to know what our biases are and how they may influence our research outcomes [27]. Thus, a variety of naturalistic data-gathering techniques will be deployed. Some of the cues that we will be questing for are repetitive words/suggestions, reflections/suggestions that may attract readers' attention (e.g., surprising ideas, unique observations, etc.), anything deemed important by the interviewee, etc.

#### 4.2.1 Inductive research approach (bottom-up approach)

In this study, we opted to follow the inductive research approach as we were dealing with unstructured data type (i.e., interviews). This bottom-up approach lets ideas, concepts and themes emerge from the interviews data.

The process contained the following steps (process flow shown in Figure. 1):

Step 1 : Data cleaning and semantic segmentation (quotes)

Step 2 : Open coding (summarization of each segment and constructing salient codes)

Step 3 : Closed coding (identifying overarching categories that can gather the open-codes)

Step 4 : Themes composition (reflective, exhaustive, relevant to the data being explored) they need to be ranked

Step 5 : Theme Analysis (examining each theme for ideas and relationship to other themes)

Step 6 : Repeat 1-5 while having constant comparison

Step 7 : Concluding remarks (common recommendations, narrative from themes, relationship between themes, own reflection)

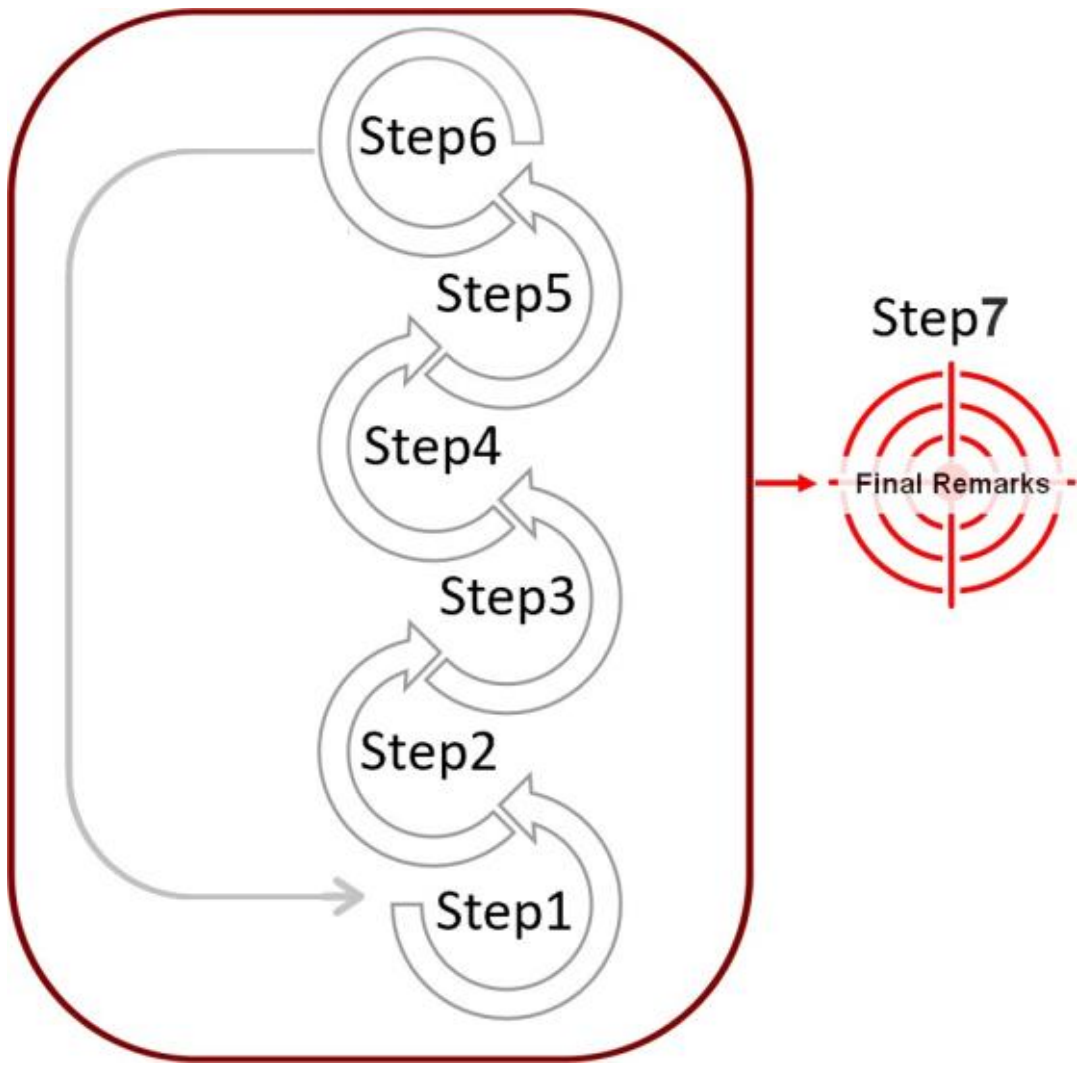

Figure 1: The inductive research approach and its iterative nature. The different steps are explained in the main text.

### 4.3 Data

Interviews were conducted with our peers, where the aim was to describe their experience with student engagement within their courses and whether they performed any adjustments. Each subject was sent a set of questions in advance for some preparation. All interviews were conducted over Zoom, due to current restrictions, and was recorded to ease transcription. The transcription was performed with the software (*Descript*[1]) and manually overlooked for correctness. All produced transcripts were included in the analysis and all recordings were deleted to address any privacy concerns.

## 4.4 Interview Discussion Points

Benefiting from the related literature review we conducted, the following contemplation set was conceived, which eventually drove the discussion with our teachers. Contemplation 1 relates to RQ1, contemplation 2-4 relate to RQ2 and contemplation 5 relates to RQ3.

**Contemplation 1 - Adaptation:** The credibility of quality education and good learning practice must be safeguarded. While it is not feasible to control and gauge students' motivation, we are accountable, as teachers, for managing their learning experience preferably espoused with enhancement of their learning engagement. In the realm of the current pandemic, distance learning activities must be adapted so that a high quality is achieved in education [29], making home a study-friendly environment, what are/could be the major adaptations which you adopted or wish to adopt?

**Contemplation 2 - Sessions Recording:** It has been proven harder to both transition from a traditional set of classroom lectures into an online format as well as monitoring the progress of students [21]. Students perceived positively the new nature of e-learning partly due to it being in the asynchronous mode, according to literature [22]. Now they are able to access the information in their own time (recorded videos of the lecture and lab sessions), instead of being forced to attend lectures etc. However, some teachers oppose this idea for it was not used during campus-based learning and in their view, it might not support student engagement (worst case the teacher will end up talking to herself/himself). What is your take on this issue?

**Contemplation 3 - "Zoom fatigue":** This is a new emerging term. Wiederhold [24] published a short editorial communication letter in which she discusses what it becomes known as the "*Zoom fatigue*". She brings about the mental exertion, lack of nonverbal cues, whether others on the Zoom call are with the teacher and following, whether the message went through, etc. Video calls leave students and teachers alike feeling unsettled and exhausted. Do you agree with this note? Why or why not? And if you do agree, what would be the remedy for this situation?

**Contemplation 4 - Virtual Reality:** To increase students' engagement and to enhance nonverbal communication, a suggestion is brought forward by Brenda [24] to shift from Zoom to other platforms that allow learning through VR with avatars (such as the *Spatial* [2] platform). "Zoom is not a good replacement for being in the office with other people, whereas something like VR gives you that level of presence and personification." Apart from the hardware constraint (special VR headsets), what do you think of this proposition (e.g., pros/cons, possibility to increase student engagement, etc.)?

[1]https://www.descript.com/

[2]Spatial: is a start-up company that recently released a program that enables people to meet through augmented reality or VR technology.

**Contemplation 5 - Gained Experience:** During the pandemic a lot of compromises, challenges and changes occurred in the courses we conducted. Some studies show that during the pandemic, students' performance and satisfaction increased. What is the lesson(s) learned and how do you envision benefiting from that when education returns to normal (e.g., implementing blended/hybrid learning)?

# 5 Results

## 5.1 Transcription Coding and Keyword Extraction

Word cloud is a digital approach to help spot frequent relevant keywords in large texts. These highlighted keywords will pin-point their importance. Prior to generating the word cloud, we cleaned the transcribed text. For instance, we removed all pronouns, adverbs, prepositions, and conjunction interjections. We then aggregated all texts pertaining to interviewees' answer to each contemplation. The overall result of the word cloud visualisation is shown in Figure 2.

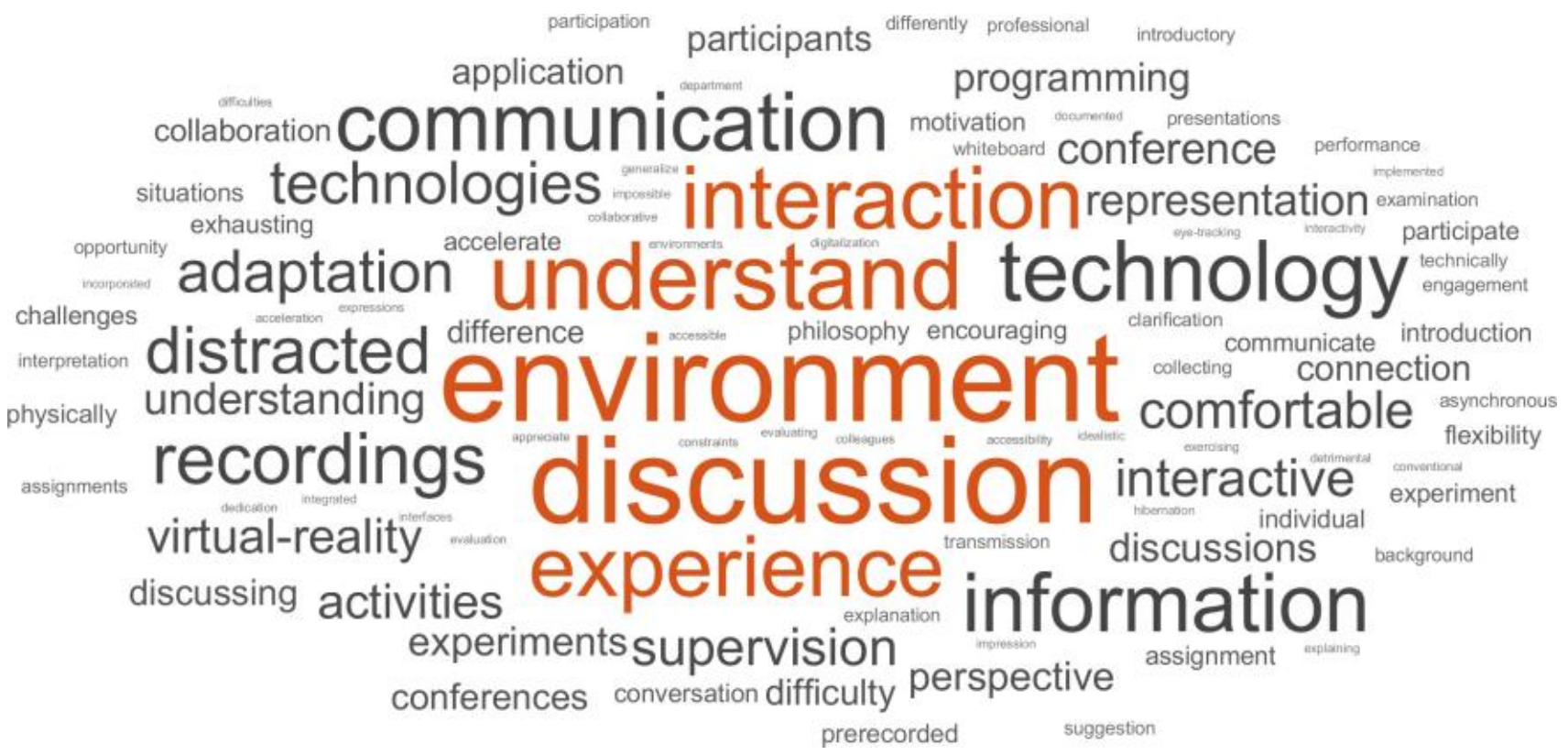

Figure 2: Word cloud chart from transcribed text pertaining to contemplation 1-5 listed in Section 4.4

## 5.2 Construction of Themes

As mentioned earlier, we adopted the inductive research approach. After all interview sessions were transcribed, the authors independently examined the text by performing the steps reported in Section 4.2.1. Then, to validate our findings, we discussed and merged the common themes and their underlying entities (see Figure 3).

*Interactivity*: One of the most apparent themes that appeared during the interviews was interaction. All teachers mentioned that the interaction with students was the significant change that occurred during the distance teaching episode. At the beginning of the pandemic, most students did not participate verbally during lectures, e.g., asking questions. Therefore, it was hard to use the chat proficiently for the teachers, making it harder to gauge their understanding. One teacher mentioned that sometimes you just had to assume that the topic was understood. However, as time progressed and the more comfortable they became with the technologies, including Zoom's polls and chat, the more interaction they received from the students.

This was especially important for their lectures. One teacher specifically mentioned that using asynchronous communication during the lectures was an important step to improve the interaction with students. Having the students write their questions in a private chat with the teacher removed the shyness barrier that can be present for some students, primarily when teaching large classes. Most of the lectures produced by the interviewees were broadcasted live and then often recorded for convenience. To help gauge the students' knowledge and understanding of the lectured topic, the polls and questions from students in the chat helped the teachers in the lectures. Two teachers brought up the topic of students learning from each other. One teacher perceived it as a negative aspect of the distance learning we provided during this pandemic. It was harder for students to interact with each other at the same level as they did in physical labs. On the contrary, another teacher mentioned a "healthy" interaction between the students in the chat during the lectures. A few who had not understood part of the topic got help from other students who grasped the topic more quickly. There were also concerns regarding formal meetings where meeting room discussions were harder to achieve, as whiteboards and related tools were not feasible to deploy.

*Availability*: Another common aspect we gathered was the theme of availability during the pandemic. In contrast to campus-based teaching, it was generally not seen as possible to provide recordings of presented lectures, increasing the course's costs and labour for the teachers. However, when the shift to distance teaching occurred, this old hassle vanished. It is immensely easier in this education setting to provide recorded lectures than it used to be. All teachers but one agreed that their courses should offer recorded lectures, both for those who intend to re-watch it later on in the course and those not attending or having network problem. Some interviewees mentioned that smaller video segments, e.g., specific videos that focus on one specific area, would also benefit the courses. However, concerns were raised regarding students not attending the live lectures and only viewing the recorded lectures. The primary concern about this scenario was that the students would not have the same opportunities to ask questions or ask for clarifications during the lecture. It may also promote procrastination for some students. However, they all agree that lectures in a recorded format were needed and could lead to a higher understanding of topics, as the students can revisit the lectures.

*Digital Tools*: One of the most significant challenges for the teachers was the sudden need for digital tools to conduct their courses, along with all their associated problems. Most of the interviewees showed a keenness to learn the tools and started incorporating quizzes, polls, mentimeters, etc., into their lectures. After getting past the first hurdle with the technology, it was evident that the teachers wanted to adapt and learn these new tools. The majority of them also found them useful for future applications in their campus-based teaching. Furthermore, one of the teachers raised the issue of complying with the national guidelines of recordings in a governmental body. All recordings must include subtitles that suffice for those with hearing impairments if it is deemed necessary[3]. Depending on the size, students, and contents of the course, the teachers who conducted the lecture might be forced to transcribe their lectures, resulting in a significantly higher workload for each lecture.

*Strains*: With all-new adaptations, there are always some strains that come along. When the sudden shift to distance teaching happened, there were many new elements all teachers had to learn. Recording and editing videos, adapting assignments, creating home exams, making sure the exams are legally correct, etc., were all time-consuming tasks, especially in the beginning. The interviewees also brought up the increased fatigue one may experience in digital environments. Some teachers noticed that towards the end of their digital lectures, the num-ber of attending students started to drop. Furthermore, the teachers who taught students from abroad noticed their lack of reliable Internet connections, making presentations and seminars harder to conduct.

[3]Accessibility for Digital Public Services Act (2019:995). Available in Swedish: http://rkrattsbaser.gov.se/sfst?bet=2018:1937

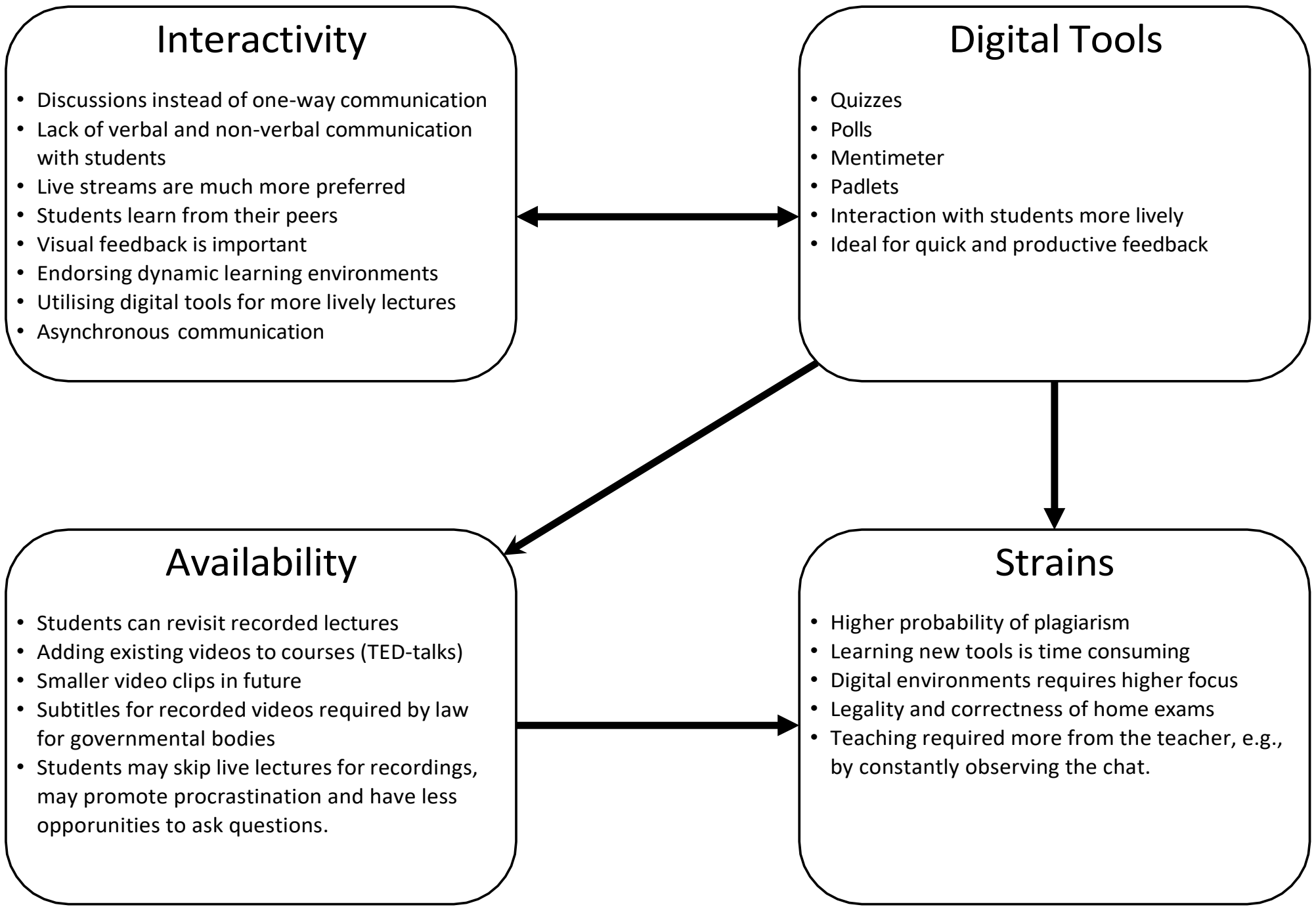

Figure 3: Themes constructed from interviews. The arrows denote the consequence/influence effect.

## 5.3 Interesting Quotes

In what follows, we gathered some quotes from the verbal interviewees' responses that we feel are interesting and need to be shared with the readers.

There was an apparent prevalence of the sense of longing for ordinary campus-based teaching among teachers. However, this pandemic impact on academic teaching setup is a temporary impetus. One teacher describes the distance teaching as: "*You [We] were doing what we could to simulate a situation there [on the Zoom platform] as if we were always at BTH.*". Another teacher added about the importance of non-verbal communication by saying: *"You get the impression that they [students] listen, because you see that they nod or they're not falling asleep at least [on campus] ... And that is what we are lacking here [distance teaching]." "...I currently feel a little bit like we're a radio station."*. According to one interviewee, the current situation is an urgent response contingent to the pandemic, and the traditional campus-based teaching shall retain its value: *"Why do you need a university [campus-based education]? You need university because of the exchange [of knowledge]. University is collaboration with students, with researchers, with companies. It's the interaction and the discussion that are interesting."*. Our teachers' sudden unprecedented exposure to contemporary technology-heavy online teaching is deemed essential to help fortify campus-based learning by borrowing new learning styles (e.g., chat) and tools. *"It seemed that maybe they [students] feel a lot more comfortable actually with us using chat."*

*"... Some things like thesis courses, some supervision, sometimes I think we will keep [online]... I think we will bring more digital tools from this process that we have started to use more."*.

A critical aspect brought up by one interviewee is the technical readiness and the pre-training. Although the practical details are different for every teacher, some teachers initially fumbled their way through a vague path that created undue strain, which was eventually surpassed by their remarkable aptitude for learning while teaching. *"We didn't, or most of us didn't have experience with providing digital courses. We are not honestly there yet on the technical readiness level and even on the application readiness level."*.

## 5.4 Answering Research Questions

In this sub-section, we try to answer our research questions which we set up in section 2.1.

**RQ1: How did our teachers adapt to distance teaching within computer science at this time of pandemic and lock-down?** All interviewees agree that adaptation was highly needed during the pandemic, e.g., in the way we interact with the students, in the way we conduct home-based exams, etc. One teacher believed that "as much as possible, we need to keep the same setting as is in campus-based teaching. The current situation mandates moving to a distance teaching, but what we are teaching is not originally distance lectures." Making ourselves acquainted with digital tools was another important aspect that came up in their interviews. Recording live lectures or using pre-recorded material was a common adaptation that all but one teacher had performed. Their reasoning for providing these recordings was that not all students function similarly. Some students might need multiple video views to fully understand the content or view the lecture in a time better suited to their daily routines. However, one concern that was repeatedly brought up regarding recorded lectures was the lack of interaction with some students. The students that only watch the recordings and never attend the live lectures might end up with questions that could have been otherwise answered during the live lecture.

**RQ2: What are possible ways that our teachers perceive to get the learning platform engaging?** During the interviews, all teachers acknowledged that it is infrequent that students turn on their webcams on Zoom during lectures. Such reaction from students made it harder for the teachers to view visual cues and gauge the students' understanding of the lectured topics and the interaction between the teachers and students seemed harder. However, using the chat within Zoom and other tools, such as Kahoot [4], polls, and Padlet [5], allowed the students to interact with the teachers asynchronously. Some teachers felt there was even more interaction than it was on campus-based courses by allowing the students to interact with the teacher asynchronously, via text or polls, which eventually removed the "shyness barrier" (according to these teachers).

**RQ3: Do our teachers believe that their new teaching experience (distance teaching) could benefit their campus-based teaching in the future?** One clear idea that all teachers were keen on bringing with them for the future was adding communication tools, such as Kahoot and polls. The asynchronous nature of interacting with the students was believed to increase the communication during lectures and made it easier to grasp whether they understood the topic. For example, small online quizzes during lectures were shown to help one of the interviewees to gauge the students' understanding. In addition, some have expressed an interest in using online lectures or recorded lectures to provide more time for seminars and other personal activities with students, leading perhaps to an improvement over the educational system with traditional lectures.

[4]https://kahoot.it/

[5]https://padlet.com/

## 6 Discussion and Recommendations

*Collective shared views:* All interviewees agree that adaptation is highly needed (e.g., in the way we interact with students, in the way we conduct the home-based exam, etc.). Making ourselves acquainted with digital tools was another important aspect that came up in their interviews. Our teachers also agree that a webcam should be activated but cannot enforce that on students. A teacher believed that "as much as possible, we need to keep the same settings as is in campus-based learning. The current situation mandates moving to distance teaching, but what we are teaching is not originally distance lectures." Another teacher mentioned that teachers workload and their attention to small details would eventually increase if we adopt the asynchronous mode. For example, remembering to record a lecture, remembering to have pause and resume, post-processing of a video and uploading and updating it (in case corrections or modifications are to be made) is time-consuming". Students may be unable to understand the whole lecture from recordings, and they must know that there is much more beyond merely watching these recordings; moreover, recording lectures do not allow interactivity (e.g., asking questions)". "Zoom fatigue" is acknowledged as a substantial effect by all teachers. This pandemic will change our teaching culture forever. Having flipped classrooms (i.e., hybrid mode) is a joint view of the teachers, where campus-based and distance-based teaching are mixed in post-pandemic education. Adopting digital tools which we got exposed to during this pandemic would be beneficial in campus-based teaching. Finally, using virtual reality and avatars to engage students in an immersive virtual classroom environment for teaching is believed to be a far-fetched and overkill solution.

*Recommendations:* The technical readiness level must be improved. We need to be better prepared for the emergency shift to distance education at any time (this was also recommended in the literature [30]). Adopting technology and digital tools (e.g., "Mentimeter", "Padlets" & "Zoom" Polls, "Kahoot" or the "Discord" platform which is believed to be much suited for interactivity) to help engage students and increase the chance to get their feedback is one of the recommendations we got from our teachers. Despite changing academic teaching approaches appears to be especially tricky, a flipped classroom where online teaching is occasionally adopted would be a good option. According to Brownell et al. [31], the didactic modelled in traditional classroom settings might not be the best way to engage large numbers of diverse populations of undergraduates.

## 7 Conclusions

Higher education took a drastic turn towards remote learning due to the procedures put in place to counteract the COVID-19 pandemic. A massive amount of effort was needed to enable the transition to a functioning remote learning structure in a concise amount of time. With more than a year of performing remote learning, this paper investigated common difficulties and remedies with the help of interviews with six of our peers (teachers in the Computer Science department). Active participation, engagement, and interaction were identified as more complex to achieve than in campus-based settings. However, the endorsement of traditional classroom education with asynchronous communication, such as chats and polls, would allow for the same, if not higher, interaction degree with the students compared to when having pure campus-based teaching. Furthermore, the change to remote learning has sparked teachers' interest in adopting such technologies to campus-based teaching to increase students' understanding and engagement. This goes hand-in-hand with the activity-based learning (ABL) concept, in which students engage in the process of learning through activities and/or discus-sions, as opposed to passively listening to an expert is deemed extremely important [32]. Thus the notion of flipped classrooms would allow integration between pedagogy and technology leading to cultivating students' learning experience [33]. We also identified mutual interest in our teachers to keep some online elements to free-up more time for face-to-face sessions with students which further corroborates the student-centered learning. Perhaps one of the noticeable pandemic bonuses is the massive crowd access to online learning environment and its associated digital tools which refined our experience and opened a new dimension to our academic life. As attributed to Winston Churchill, we should "never waste a good crisis" experience.

# Biography

**ABBAS CHEDDAD (Senior Member, IEEE) received the Ph.D. degree (Hons.) from the Univer- sity of Ulster, U.K., in 2010. He has held research positions at several universities in Sweden, such as, Umeå University and Karolinska Institute, where he focused his research on medical image analysis (disease risk prediction and optical projection tomography). He is leading a research group on big data analytics for image processing. Currently, the group is collaborating, research- wise, with three companies, SONY Mobile Communications AB, Lund, Arkiv Digital AB, Mariestad, Sweden, and GKN Aerospace AB (the world’s leading multi-technology tier 1 aerospace supplier) by addressing practical industrial problems. He is an Associate Professor (Docent) with the Blekinge Institute of Technology (BTH), Sweden. He has in records, one book, one book chapter (invited), two granted patents, and more than 60 journal articles and conference papers. He is a member of the IEEE Signal Processing Society and an ACM Distinguished Speaker. He received several awards, including the 25K Award for New Entrepreneurs in the hi-tech category sponsored by Northern Ireland Science Park (NISP) and has acquired grants that total up to 400KEUR. He was the Chair of three international conferences/workshops, a PC member in dozens of conferences, and has been invited for talks at several venues**

**Christian Nordahl is a Ph.D. Student in Computer Science at the department of Computer Science at Blekinge Institute of Technology. Christian is working in the BigData@BTH project. Current research interests include behavior modeling, data mining, clustering, and machine learning.**